\documentclass[a4paper,12pt]{article}
\usepackage{eurosym}
\usepackage[dvips]{graphicx}
\usepackage{amsfonts}
\usepackage{mathrsfs}
\usepackage{graphicx} 
\usepackage{epsfig}
\usepackage{amsmath, amsthm}
\usepackage{amssymb}

\newcommand{\ba}{\begin{array}}
\newcommand{\ea}{\end{array}}

\newcommand{\beq}{\begin{equation}}
\newcommand{\eeq}{\end{equation}}
\newcommand{\ben}{\begin{enumerate}}
\newcommand{\een}{\end{enumerate}}
\newcommand{\bit}{\begin{itemize}}
\newcommand{\eit}{\end{itemize}}

\begin{document}
\title{Notes on Integrable Motion of Two Interacting  Curves and  Two-layer Generalized Heisenberg Ferromagnet Equations}
\author{Zhaidary  Myrzakulova\footnote{Email: zhrmyrzakulov@gmail.com}, Akbota Myrzakul\footnote{Email: akbota.myrzakul@gmail.com},  Gulgassyl  Nugmanova\footnote{Email: nugmanovagn@gmail.com}\\
and Ratbay Myrzakulov\footnote{Email: rmyrzakulov@gmail.com}\\
\textsl{Eurasian International Center for Theoretical Physics and} \\ { Department of General \& Theoretical Physics, Eurasian}  \\ National University,
Astana, 010008, Kazakhstan
}
\maketitle

\begin{abstract}
In the present paper, we study the  integrable 2-layer generalized Heisenberg ferromagnet equation (HFE).  The relation between this generalized HFE  and  differential geometry of curves is established. Using this relation we found the geometrical equivalent counterpart of the 2-layer spin system which is the 2-component KdV equation. Finally, the gauge  equivalence between these equations is established. 
\end{abstract}


\section{Introduction}
The  Heisenberg ferromagnet  equation (HFE)
\begin{equation}
{\bf A}_{t}={\bf A}\wedge {\bf A}_{xx},  \label{1}
\end{equation}
is one of basic fundamental nonlinear differential equations integrable by inverse scattering transform (IST) method \cite{R13} (see also Refs. \cite{myrzakulov-391}-\cite{myrzakulov-314} and references therein). Above,  ${\bf A}=(A_{1},A_{2},A_{3})$ is a unit spin vector. As well-known, the HFE is geometrically and gauge equivalent to the nonlinear Schrodinger equation (NLSE)
\begin{equation}
iq_{t}+q_{xx}+2|q|^{2}q=0.  \label{2}
\end{equation}
The HFE   describes the nonlinear dynamics of 1-layer ferromagnets \cite{Chen1}-\cite{myrzakulov-1397}. At the same, it is well-known that ferromagnets have the multilayer nature. To describe such multilayer ferromagnets we need some type multilayer generalizations of the HFE (1) \cite{M35}. This is the first and main physical motivation of this paper.  On the other hand, there are a lot of multicomponent generalizations of known integrable systems. For example, the NLSE (2) admits the following integrable 2-component generalization
\begin{eqnarray}
iq_{1t}+q_{1xx}+2(|q_{1}|^{2}+|q_{2}|^{2})q_{1}&=&0,  \label{3}\\
iq_{2t}+q_{2xx}+2(|q_{1}|^{2}+|q_{2}|^{2})q_{2}&=&0,  \label{4}
\end{eqnarray}
which is known as the Manakov system.  Such type integrable generalizations of soliton equations dictates also needs to construct integrable and nonintegrable multilayer extensions of the HFE (1). It is the second (may be mathematical) motivation  for our study of multilayer generalizations of the HFE (1). Note that the one of 2-layer spin systems was constructed in \cite{akb4}-\cite{akb8} and reads as
\begin{eqnarray}
{\bf A}_{t}&=&{\bf A}\wedge {\bf A}_{xx}+u_{1}{\bf A}_{x}+v_{1}{\bf H}_{1}\wedge{\bf A},\\ \label{5}
{\bf B}_{t}&=&{\bf B}\wedge {\bf B}_{xx}+u_{2}{\bf B}_{x}+v_{2}{\bf H}_{2}\wedge{\bf B},  \label{6}
\end{eqnarray}
where ${\bf B}=(B_{1}, B_{2}, B_{3})$ is the second spin vector and $u_{j}, v_{j}$ are some real functions of $A_{j}, B_{j}$. This set of equations (which known as the 2-layer Myrzakulov -LIII (M-LIII) equation) is the gauge and geometrical equivalent counterpart of the Manakov system (3)-(4) and hence is integrable.

The outline of the present paper is organized as follows. In section 2, we
present  the 2-layer M-IV  equation. Also the relation between the motion of space curves and the 2-layer M-IV  equation is established. Then using this relation we found that the Lakshmanan (geometrical) equivalent counterpart of the M-LXXIII equation is the well-known 2-component KdV  equation. Section 3 is devoted to the integrable aspects of the considered systems. The gauge equivalence between the 2-layer M-IV  equation and the 2-component KdV  equation is established  in Section 4. Some briefly information for the surface induced by the $SU(3)$ $\Gamma$ - spin system is presented in Section 5. The paper is concluded by some comments  in
Section 6.

\section{Integrable 2-layer spin system}

In this paper, we consider the following 2-layer spin system called the 2-layer Myrzakulov-IV (M-IV) equation
\begin{eqnarray}
{\bf A}_{t}&=&{\bf A}_{xxx}+u_{1}{\bf A}_{x}+v_{1}{\bf A},\\ \label{2}
{\bf B}_{t}&=&{\bf B}_{xxx}+u_{2}{\bf B}_{x}+v_{2}{\bf B},  \label{2}
\end{eqnarray}
where ${\bf A}=(A_{1}, A_{2}, A_{3}), \quad {\bf B}=(B_{1}, B_{2}, B_{3})$ are unit  spin vectors $({\bf A}^{2}={\bf B}^{2}=1)$ and 
 \begin{eqnarray}
u_{1}&=& {\bf A}_{x}^{2}+3(\sqrt{{\bf A}_{x}^{2}}+\sqrt{{\bf B}_{x}^{2}}), \quad v_{1}=\frac{3}{2}({\bf A}_{x}^{2})_{x},\\
u_{2}&=& {\bf B}_{x}^{2}+3(\sqrt{{\bf A}_{x}^{2}}+\sqrt{{\bf B}_{x}^{2}}), \quad v_{2}=\frac{3}{2}({\bf B}_{x}^{2})_{x}.
\end{eqnarray}

We will study  this set of equations from the different point of views. We start from the construction the relation between the 2-layer M-IV equation  (7)-(8) and differential  geometry of curves. Consider two interacting space curves in $R^{4}$  space. With each curves  is related two set of unit vectors ${\bf e}_{j}$ and ${\bf l}_{j}$, respectively. They satisfy the Frenet - Serret equations. Before  going further, here it is appropriate to make a few comments that is here some algebraic comments in order. In the level of Lie algebras we have that algebra so(4) matrices  are just antisymmetric matrices $4\times 4$. It turns out that the six-dimensional space of such $4\times 4$ matrices decomposes into two three-dimensional subspaces that are each closed under taking commutators and each of them satisfies precisely the commutation relations of $so(3)$.  So here we use just the well-known algebraic relation $so(4)=so(3)\times so(3)$ and/or for space  $R^{4}=R^{3}\times R^{3}$. 
  
It is well-known that between some integrable (soliton) equations take place the so-called geometrical (Lakshmanan) equivalence. In this section, our aim is to  find the Lakshmanan equivalent counterpart of the    equation (7)-(8).  To do that, let us return to the interacting two  3-dimensional curves  in  $R^{4}$. The motion of these curves is  given by  the following equations 
\begin{eqnarray}
\left ( \begin{array}{ccc}
{\bf  e}_{1} \\
{\bf  e}_{2} \\
{\bf  e}_{3}
\end{array} \right)_{x} &=& C_{1}
\left ( \begin{array}{ccc}
{\bf  e}_{1} \\
{\bf  e}_{2} \\
{\bf  e}_{3}
\end{array} \right),\quad
\left ( \begin{array}{ccc}
{\bf  e}_{1} \\
{\bf  e}_{2} \\
{\bf  e}_{3}
\end{array} \right)_{t} =G_{1}
\left ( \begin{array}{ccc}
{\bf  e}_{1} \\
{\bf  e}_{2} \\
{\bf  e}_{3}
\end{array} \right), \\ 
\left ( \begin{array}{ccc}
{\bf  l}_{1} \\
{\bf  l}_{2} \\
{\bf  l}_{3}
\end{array} \right)_{x} &=& C_{2}
\left ( \begin{array}{ccc}
{\bf  l}_{1} \\
{\bf  l}_{2} \\
{\bf  l}_{3}
\end{array} \right),\quad
\left ( \begin{array}{ccc}
{\bf  l}_{1} \\
{\bf  l}_{2} \\
{\bf  l}_{3}
\end{array} \right)_{t} = G_{2}
\left ( \begin{array}{ccc}
{\bf  l}_{1} \\
{\bf  l}_{2} \\
{\bf  l}_{3}
\end{array} \right).  
\end{eqnarray}
Here ${\bf e}_{j}$ and ${\bf l}_{j}$  are the unit tangent ($j=1$), normal ($j=2$)
and binormal ($j=3$) vectors  to the   curves,  $x$ is its common arclength 
parametrising the curves. The matrices $C_{j}$ and $G_{j}$ have the forms
\begin{eqnarray}
C_{1} =
\left ( \begin{array}{ccc}
0   & \kappa_{1}     & 0 \\
-\kappa_{1} & 0     & \tau  \\
0   & -\tau & 0
\end{array} \right) ,\quad
G_{1} =
\left ( \begin{array}{ccc}
0       & \omega_{3}  & -\omega_{2} \\
-\omega_{3} & 0      & \omega_{1} \\
\omega_{2}  & -\omega_{1} & 0
\end{array} \right),\label{2.2} 
\end{eqnarray}
\begin{eqnarray}
C_{2} =
\left ( \begin{array}{ccc}
0   & \kappa_{2}     & 0 \\
-\kappa_{2} & 0     & \tau_{2} \\
0   & -\tau_{2} & 0
\end{array} \right),\quad
G_{2} =
\left ( \begin{array}{ccc}
0       & \theta_{3}  & -\theta_{2} \\
-\theta_{3} & 0      & \theta_{1} \\
\theta_{2}  & -\theta_{1} & 0
\end{array} \right).\label{2.2} 
\end{eqnarray}
Curvatures and torsions of   curves  are given  by the following formulas
\begin{eqnarray}
\kappa_{1}&=& \sqrt{{\bf e}_{1x}^{2}},\quad \tau_{1}= \frac{{\bf e}_{1}\cdot ({\bf e}_{1x} \wedge {\bf e}_{1xx})}{{\bf e}_{1x}^{2}},        \\
\kappa_{2}&=& \sqrt{{\bf l}_{1x}^{2}},\quad \tau_{2}= \frac{{\bf l}_{1}\cdot ({\bf l}_{1x} \wedge {\bf l}_{1xx})}{{\bf l}_{1x}^{2}}.        
\end{eqnarray}
The compatibility condition of the equations (11)-(12) reads as 
\begin{eqnarray}
C_{1t} - G_{1x} + [C_{1}, G_{1}] &=& 0,\\
C_{2t} - G_{2x} + [C_{2}, G_{2}] &=& 0 
\end{eqnarray}
or in elements   
 \begin{eqnarray}
\kappa_{1t}    & = & \omega_{3x} + \tau_{1} \omega_2, \label{2.5} \\ 
\tau_{1t}      & = & \omega_{1x} - \kappa_{1}\omega_2, \\ \label{2.6} 
\omega_{2x} & = & \tau \omega_3-\kappa \omega_1, \label{2.7} \\
\kappa_{2t}    & = & \theta_{3x} + \tau_{1} \theta_2, \label{2.5} \\ 
\tau_{2t}      & = & \theta_{1x} - \kappa_{1}\theta_2, \\ \label{2.6} 
\theta_{2x} & = & \tau_{2} \theta_3-\kappa \theta_1. \label{2.7} 
\end{eqnarray}

Now we do the following identifications:
 \begin{eqnarray}
{\bf A}\equiv {\bf e}_{1}, \quad  {\bf B}\equiv {\bf l}_{1}. \label{2.7} 
\end{eqnarray}
Then we have 
\begin{eqnarray}
\kappa^{2}_{1} & = & {\bf A}_{x}^{2},\quad \tau_{1}=  \frac{{\bf A}\cdot ({\bf A}_{x} \wedge {\bf A}_{xx})}{{\bf A}_{x}^{2}}, \\
\kappa^{2}_{2} & = & {\bf B}_{x}^{2},\quad \tau_{2}=  \frac{{\bf B}\cdot ({\bf B}_{x} \wedge {\bf B}_{xx})}{{\bf B}_{x}^{2}}. 
\end{eqnarray}
Now we want to simplify the problem. Namely we assume that
\begin{eqnarray}
 \tau_{1}=  \tau_{2}= 0. 
\end{eqnarray}
This means we consider two  plane curves without torsions. In this case we have
\begin{eqnarray}
\omega_{2}  = \omega_{1}=\theta_{1} =\theta_{3}=0.       
\end{eqnarray}
At the same time from (19)-(24) we get
 \begin{eqnarray}
\kappa_{1t}    = \omega_{3x}, \quad \kappa_{2t}    = \theta_{3x}. \label{30} 
\end{eqnarray}
These results give us
\begin{eqnarray}
\omega_{3} & = & -\kappa_{1xx}-3(\kappa_{1}+\kappa_{2})\kappa_{1}, \\
\theta_{3} & = & -\kappa_{2xx}-3(\kappa_{1}+\kappa_{2})\kappa_{2}.        
\end{eqnarray}
Finally we obtain the following set of equations
 \begin{eqnarray}
\kappa_{1t}    +\kappa_{1xxx}+3[(\kappa_{1}+\kappa_{2})\kappa_{1}]_{x}&=&0, \\ 
\kappa_{2t}   +\kappa_{2xxx}+3[(\kappa_{1}+\kappa_{2})\kappa_{2}]_{x}&=&0.
\end{eqnarray}
It is the desired  set of equations and which  is the Lakshmanan or geometrical equivalent counterpart of the 2-layer M-IV equation  (7)-(8). Integrability properties of the obtained set (33)-(34) and the 2-layer M-IV equation (7)-(8) we will consider in the next sections.

\section{Integrability of the 2-layer M-IV equation}

Our aim in this section is to study integrability aspects of the M-IV equation (7)-(8). To this end, let us consider the Lax representation of the form
\begin{eqnarray}
\Psi_x&=&U\Psi,\\ 
\Psi_t&=&V\Psi, 
\end{eqnarray}
where 
\begin{eqnarray}
U&=&-i\lambda \Sigma+Q, \\
V&=&-4i\lambda^{3}\Sigma+4\lambda^{2}Q+2i\lambda F_{1}+F_{0}.
\end{eqnarray}
Here
\begin{eqnarray}
\Sigma=\begin{pmatrix} 1&0&0\\0&-1&0\\
0&0&-1\end{pmatrix}, \quad Q=\begin{pmatrix} 0&q_{1}&q_{2}\\r_{1}&0&0\\
r_{2}&0&0\end{pmatrix}, 
\end{eqnarray}
\begin{eqnarray}
F_{1}=\begin{pmatrix} -v&q_{1x}&q_{2x}\\
-r_{1x}&r_{1}q_{1}&r_{1}q_{2}\\
-r_{2x}&r_{2}q_{1}&r_{2}q_{2}\end{pmatrix}, \quad F_{0}=-Q_{xx}-2v\Sigma Q-F_{02}, 
\end{eqnarray}
where $v=r_{1}q_{1}+r_{2}q_{2}$ and 
\begin{eqnarray}
F_{02}=\begin{pmatrix} (r_{1x}q_{1}-r_{1}q_{1x})+(r_{2x}q_{2}-r_{2}q_{2x})&0&0\\
0&r_{1}q_{1x}-r_{1x}q_{1}&r_{1}q_{2x}-r_{1x}q_{2}\\
0&r_{2}q_{1x}-r_{2x}q_{1}&r_{2}q_{2x}-r_{2x}q_{2}\end{pmatrix}.
\end{eqnarray}
The compatibility condition 
\begin{eqnarray}
U_{t}-V_{x}+[U,V]=0
\end{eqnarray}
 gives the following set of equations
 \begin{eqnarray}
q_{1t}    +q_{1xxx}-3r_{1}(q_{1}^{2})_{x}-3r_{2}(q_{1}q_{2})_{x}&=&0, \\ 
q_{2t}    +q_{2xxx}-3r_{2}(q_{2}^{2})_{x}-3r_{1}(q_{1}q_{2})_{x}&=&0,\\
r_{1t}    +r_{1xxx}-3q_{1}(r_{1}^{2})_{x}-3q_{2}(r_{1}r_{2})_{x}&=&0,\\
r_{2t}    +r_{2xxx}-3q_{2}(r_{2}^{2})_{x}-3q_{1}(r_{1}r_{2})_{x}&=&0.
\end{eqnarray}
 It can be considered as the 2-component modified Korteweg-de Vries (KdV) equation. We now present some  reductions of the equations (43)-(46). 

i) Let $r_{1}=r_{2}=-1$. In this case the set of equations (43)-(46) takes the form
 \begin{eqnarray}
q_{1t}    +q_{1xxx}+3[(q_{1}+q_{2})q_{1}]_{x}&=&0, \\ 
q_{2t}    +q_{2xxx}+3[(q_{1}+q_{2})q_{2}]_{x}&=&0.
\end{eqnarray}
It is the 2-component KdV equation, which is integrable as the exact reduction of the integrable set of equations (43)-(46).

ii) Now we consider the case $r_{1}=\sigma_{1}q_{1}, \quad r_{2}=\sigma_{2}q_{2}$, where $\sigma_{j}=\pm 1$. For this reduction, the set (43)-(46) converted to the following set of equations
\begin{eqnarray}
q_{1t}    +q_{1xxx}-6\sigma_{1}q_{1}^{2}q_{1x}-3\sigma_{2}q_{2}(q_{1}q_{2})_{x}&=&0, \\ 
q_{2t}    +q_{2xxx}-6\sigma_{2}q_{2}^{2}q_{2x}-3\sigma_{1}q_{1}(q_{1}q_{2})_{x}&=&0.
\end{eqnarray}
It is nothing but the integrable 2-component mKdV equation.

iii) Our third  example is the following reduction: $r_{1}=\sigma_{1}{\bar q}_{1}, \quad r_{2}=\sigma_{2}{\bar q}_{2}$. In this case we have
\begin{eqnarray}
q_{1t}    +q_{1xxx}-6\sigma_{1}{\bar q}_{1}(q_{1}^{2})_{x}-3\sigma_{2}{\bar q}_{2}(q_{1}q_{2})_{x}&=&0, \\ 
q_{2t}    +q_{2xxx}-6\sigma_{2}{\bar q}_{2}(q_{2}^{2})_{x}-3\sigma_{1}{\bar q}_{1}(q_{1}q_{2})_{x}&=&0,
\end{eqnarray}
which is the well-known 2-component complex mKdV equation.

iv) Our last example is following: $r_{1}=\sigma_{1}q_{1}, \quad r_{2}=\sigma_{2}{\bar q}_{2}$, where we assume that $q_{1}$ is real function and $q_{2}$ is complex function. Then the equations (43)-(46) take the form
\begin{eqnarray}
q_{1t}    +q_{1xxx}-6\sigma_{1}q_{1}^{2}q_{1x}-3\sigma_{2}{\bar q}_{2}(q_{1}q_{2})_{x}&=&0, \\ 
q_{2t}    +q_{2xxx}-3\sigma_{2}{\bar q}_{2}(q_{2}^{2})_{x}-3\sigma_{1}q_{1}(q_{1}q_{2})_{x}&=&0.
\end{eqnarray}
It is some kind integrable set of the mixed mKdV-complex mKdV  equations. From results of this section follows that the equivalent counterpart of the 2-layer M-IV equation (7)-(8)  is the 2-component KdV equation (47)-(48) which has the same form with (33)-(34).

\section{Gauge equivalent equation}

It is interesting to find the gauge equivalent  equation to the set of equations (7)-(8) that is to the 2-layer M-IV equation. With this aim, let us we consider the gauge transformation $\Phi=g^{-1}\Psi$, where $g=\Psi|_{\lambda=0}$. Then we have
\begin{eqnarray}
\Phi_x&=&-i\lambda \Gamma\Phi,\\ 
\Phi_t&=&[-4i\lambda^{3}\Gamma+2\lambda^{2}\Gamma\Gamma_{x}+i\lambda(\Gamma_{xx}+\frac{3}{2}\Gamma\Gamma^{2}_{x})]\Phi, 
\end{eqnarray}
where
\begin{eqnarray}
 \Gamma=g^{-1}\Sigma g=\begin{pmatrix}\Gamma_{11}&\Gamma_{12}&\Gamma_{13}\\
\Gamma_{21}&\Gamma_{22}&\Gamma_{23}\\
\Gamma_{31}&\Gamma_{32}&\Gamma_{33}\end{pmatrix}. 
\end{eqnarray}
The compatibility condition of the equations (55)-(56) gives
\begin{eqnarray}
\Gamma_{t}+\Gamma_{xxx}+\frac{3}{2}(\Gamma_{x}^{2}\Gamma)_{x}=0. 
\end{eqnarray}
This equation  can be considered as the $su(3)$ form of the 2-layer M-IV equation (7)-(8). From the definition of the $\Gamma$ -  matrix function follows that
\begin{eqnarray}
\Gamma_{x}=g^{-1}[\Sigma,Q]g 
\end{eqnarray}
so that
\begin{eqnarray}
\Gamma_{x}^{2}=-4\begin{pmatrix}v&0&0\\
0&r_{1}q_{1}&r_{1}q_{2}\\
0&r_{2}q_{1}&r_{2}q_{2}\end{pmatrix}.  
\end{eqnarray}
Hence we obtain
\begin{eqnarray}
tr(\Gamma^{2}_{x})=-8v. 
\end{eqnarray}
It is the Hamiltonian of the equation (58)
\begin{eqnarray}
H=\frac{1}{2}\int tr(\Gamma^{2}_{x}). 
\end{eqnarray}
Note that from (62) we can get the Hamiltonian of the 2-layer M-IV equation (7)-(8)  with the suitable reduction.

\section{Integrable surface}
Lastly let us briefly construct an integrable surface related with the  equation (58). We assume the following identification
\begin{eqnarray}
\Gamma\equiv R_{x}. 
\end{eqnarray}
Then from (58) we have
\begin{eqnarray}
R_{t}+R_{xxx}+\frac{3}{2}R_{xx}^{2}R_{x}=0. 
\end{eqnarray}
Now we can define the first fundamental form of the surface as
\begin{eqnarray}
I=tr(R^{2}_{t})dt^{2}+2tr(R_{t}R_{x})dxdt+tr(R_{x}^{2})dx^{2}.
\end{eqnarray}
Similarly we can construct the second fundamental form. These two forms allow us to construct the surface induced by the spin system (58) and which is integrable naturally.  
\section{Conclusions}
 
In this paper, we have established the relation between the 2-layer M-IV  equation  (7)-(8) and the 2-component KdV equation (47)-(48).  We have shown that the 2-layer M-IV  equation (7)-(8) and the 2-component KdV equation (47)-(48)  is the geometrically equivalent each to other. Also the gauge  equivalence between these equations is proved. 
Our results are significant for the deep understand
of integrable spin systems and their relations with differential geometry of curves and surfaces in multilayer and multicomponent case. 
Note that the 2-component KdV  equation  and the 2-component mKdV equation  as well as the set of equations (43)-(46) can be viewed as one of  members of the Manakov  system hierarchy (3)-(4). Finally we note that there are some integrable 2-layer spin systems (that is generalized HFE) in 2+1 dimensions. In particular,  in \cite{28} was constructed the 2-layer Myrzakulov-I (M-I) equation of the form
\begin{eqnarray}
{\bf A}_{t}&=&{\bf A}\wedge {\bf A}_{xy}+u_{1}{\bf A}_{x}+v_{1}{\bf H}_{1}\wedge{\bf A}+{\bf F}_{1},\\ 
{\bf B}_{t}&=&{\bf B}\wedge {\bf B}_{xy}+u_{2}{\bf B}_{x}+v_{2}{\bf H}_{2}\wedge{\bf B}+{\bf F}_{2}, \\
u_{1x}&=&-{\bf A}\cdot({\bf A}_{x}\wedge {\bf A}_{y}),\\
u_{2x}&=&-{\bf B}\cdot({\bf B}_{x}\wedge {\bf B}_{y}),
\end{eqnarray}
where $F_{j}$ are some vector functions \cite{28}-\cite{29}. In fact, it  is the integrable 2-layer M-I equation. It is well-known that the M-I equation  is one of  integrable (2+1)-dimensional generalizations of the HFE (1). The other important example integrable (2+1)-dimensional spin systems  is the famous Ishimori equation. Its integrable 2-layer generalization was presented in \cite{29} and reads as
\begin{eqnarray}
{\bf A}_{t}&=&{\bf A}\wedge ({\bf A}_{xx}+{\bf A}_{yy})+u_{1y}{\bf A}_{x}+u_{1x}{\bf A}_{y}+v_{1}{\bf H}_{1}\wedge{\bf A}+{\bf F}_{1},\\ 
{\bf B}_{t}&=&{\bf B}\wedge ({\bf B}_{xx}+{\bf B}_{yy})+u_{2y}{\bf B}_{x}+u_{2x}{\bf B}_{y}+v_{2}{\bf H}_{2}\wedge{\bf B}+{\bf F}_{2},\\
u_{1xx}&-&\alpha^{2}u_{1yy}=-2\alpha^{2}{\bf A}\cdot({\bf A}_{x}\wedge {\bf A}_{y}),\\
u_{2xx}&-&\alpha^{2}u_{2yy}=-2\alpha^{2}{\bf B}\cdot({\bf B}_{x}\wedge {\bf B}_{y}).
\end{eqnarray} 
At last we note that the construction of integrable class of two and (in general)  $n$ interacting curves and surfaces is one of actual problems of modern differential geometry of curves and surfaces.  Our work in this direction in progress. 

 \section{Acknowledgements}
This work was supported  by  the Ministry of Edication  and Science of Kazakhstan under
grants 0118РК00935 and 0118РК00693.

 \end{document}